\documentclass[aps,pra,reprint,twocolumn,amsmath,amssymb,citeautoscript,longbibliography]{revtex4-1}
\usepackage{graphicx}
\usepackage{dcolumn}
\usepackage{bm}
\usepackage{latexsym,epsfig}
\usepackage{graphicx}
\usepackage{verbatim}
\usepackage{comment}
\usepackage{amsmath}
\usepackage{amssymb}
\usepackage{physics}
\usepackage{stmaryrd}
\usepackage{color}
\usepackage{epstopdf}
\usepackage{grffile}
\usepackage{lipsum}
\usepackage{enumitem}
\usepackage{ulem}

\usepackage[dvipsnames]{xcolor}
\DeclareGraphicsExtensions{.eps}

\newcommand{\beq}{\begin{equation}}
\newcommand{\eeq}{\end{equation}}
\newcommand{\bea}{\begin{eqnarray}}
\newcommand{\eea}{\end{eqnarray}}
\newcommand{\ben}{\begin{eqnarray*}}
\newcommand{\een}{\end{eqnarray*}}
\newcommand{\bfig}{\begin{figure}}
\newcommand{\efig}{\end{figure}}

\usepackage{hyperref}
\hypersetup{
    colorlinks=true,      
    urlcolor=blue,
    citecolor=blue,
    linkcolor=blue
}

\begin{document}
\title{Non-Hermitian comb effect in coupled clean and quasiperiodic chains}

\author{Soumya Ranjan Padhi, Souvik Roy, Biswajit Paul, Sanchayan Banerjee, and Tapan Mishra}

\email{mishratapan@gmail.com}

\affiliation{School of Physical Sciences, National Institute of Science Education and Research, Jatni 752050, India}

\affiliation{Homi Bhabha National Institute, Training School Complex, Anushaktinagar, Mumbai 400094, India}

\date{\today}

\begin{abstract}
We study localization properties in a system of non-Hermitian quasiperiodic chain coupled to a uniform chain or clean chain by inter-chain hopping. 
We find that in the limit of weak inter-chain coupling, such a coupled system exhibits transitions from delocalized to intermediate phase with increase in the non-Hermiticity parameter. However, for stronger inter-chain coupling strengths, the delocalized phase undergoes a transition to localized phase and then to an intermediate phase. Interestingly, the intermediate phase in this case exhibits the non-Hermitian comb effect (NHCE), i.e., the coexistence of localized and extended states rather than being well separated from each other by any mobility edge which is conventional in any intermediate phase. We further show that such a NHCE originates from the isolated site limit of the quasiperiodic chain and provide an analytical explanation supporting the numerical signatures. 
\end{abstract}

\maketitle

\section{Introduction} 
\label{sec:intro}
The study of localization transition has been one of the most fascinating areas of research starting from the seminal work of Anderson~\cite{Anderson_1958} that suggests exponential localization of single particle states in lattices with uncorrelated (random) disorder. While in three dimension, there exists a well defined localization transition at a critical disorder strength, 
in one dimension (1D), all eigenstates are localized for arbitrarily weak random disorder in the system~\cite{Gang_of_four_1979}. Quasiperiodic (QP) potentials on the other hand, provide a notable exception to this. The one dimensional Aubry–Andr\'e (AA) model — a tight binding lattice with an onsite QP potential — exhibits a sharp delocalization to localization transition at a critical potential strength owing to the self-duality of the model~\cite{Aubry_1980,Paredesreview_2019,Roati2008}. Extending the AA model to include long-range hopping, hopping dimerization, or other generalized potentials yield intermediate phases with coexisting localized and extended states separated by the well-known mobility edges~\cite{ME_bich_sarma_2017,ME_Incomm_Opt_sarma2010,Roy_prl_2021,Pedro_2023,PRB_GME_sarma_2023,Loc_AA_non-nerest,Santos_prl_2019,Auditya_2021,Tanay_PRR_2020,Tanay_PRE_2020,ME_IP_sarma2020,ME_Shallow_1DQP_potentials,NN_TB_sarma2015,padhan_prbl_2022,Immanuel_prl_2018,Xiong_2020,subrotoreview_2017,expt_Int_ME_GAA}. 

Non-Hermitian (NH) quasiperiodic  models on the other hand exhibits striking localization effects. The NH variants of the AA model show sharp localization transitions induced either by tuning the strength of the complex quasiperiodic potential, or by introducing a complex phase in the potential~\cite{Longhi_prl_2019}. Similar phase transition can also be seen by introducing non-reciprocal hopping in AA model.
~\cite{Shuchen_2019}. 
Furthermore, $\mathcal{PT}$ symmetry breaking in such systems yields localized states with complex energies, and other generalized models host mobility edges that form closed “mobility rings’’ in the complex-energy plane~\cite{Zhang_pra_2021, Cai_2022, Shuchen_es_2021, Zhou_prb_2023, Soutang_rent_2023, Zhou_dimerization_2022, Wu_iop_2021, Zeng_prb_2020, ShuChen_EMEs_2021, ShuChen_Exp_decay_2021, Wang_unconventional_2021, Longhi_maryland_2021, Zhou_floquet_2021, Zhou_marryland_2022, Zeng_GMEs_NHGAA_expt_2020, Li_mobrings_2024, Padhan_prbl_nh_2023, Gandhi_kitaev_2024, Peng_power_nh_2023, ShuChen_power_nh_2021, Tong_liu_2020, AMES_arxiv_2024}. Many of these phenomena including ME physics, have now been observed across photonic lattices~\cite{Tong_liu_2020, Longhi_prl_2019, Lin2022}, ultracold atoms~\cite{Zhou, BoYan, Ren_2022}, electric circuits~\cite{Zeng_GMEs_NHGAA_expt_2020, Liu_circuit_2023, Zhang2023}, and periodically driven systems~\cite{Weidemann2022}.

Going beyond strictly one dimensional limit, the systems of two coupled chains provide useful platforms to uncover the effect of inter-chain couplings on the localization properties. 
A simplest yet interesting scenario in this context is the the quasiperiodic chain coupled to a clean chain (no disorder) through inter-chain hopping. The Hermitian limits of such a system exhibits delocalization to localization transition through an intermediate phase with well defined ME due to the effect of strong inter-chain coupling~\cite{ME_IP_sarma2020}. 

In this work, however, we show that when a non-Hermitian quasiperiodic chain is coupled to a clean Hermitian chain, interesting scenarios appear in the single particle spectrum. We obtain that for a fixed quasiperiodic potential strength, the non-Hermiticity parameter can induce a delocalization to intermediate phase transition in the limit of vanishing and weak inter-chain hopping. However, when the inter-chain hopping is strong, the system undergoes delocalization to localization transition through an intermediate phase. A further increase in the strength of the non-Hermiticity parameter reintroduces the intermediate phase even if the system is already in the localized phase. Interestingly, we find that for certain combinations of inter-chain hopping and non-Hermiticity parameter, the intermediate phase hosts many isolated localized states intermixed with the continuum of delocalized states - a phenomenon recently predicted as the non-Hermitian comb effect (NHCE)~\cite{COMB_2025}. Note that the intermeidiate phase that exhibits the NHCE  is fundamentally different from the conventional intermediate phases where the delocalized and localized states are separated from each other by the ME. We also obtain that the NHCE is robust against strong quasiperiodic potential strength. In the end, we explain the origin of the NHCE by drawing insights from the isolated site limits of the quasiperiodic chain.

The rest of the paper is organized as follows. In Sec.~\ref{sec:model}, we lay out the model and approach to study the properties of the coupled system. In Sec.~\ref{sec:results}, we discuss phase diagram, non-Hermitian comb effect in weak and strong inter-chain coupling. Finally, in Sec.~\ref{sec:conc}, we summarize our results and offer a brief outlook.


\begin{figure}[t]
\centering
\includegraphics[width=0.85\columnwidth]{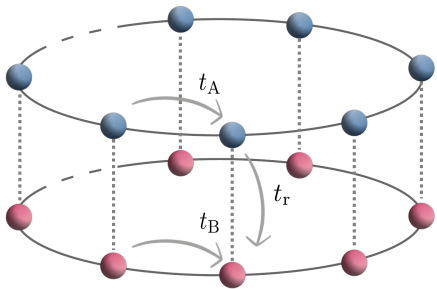}
\caption{
Schematic of a coupled non-Hermitian quasiperiodic an clean chain system. The red filled circles are sites subjected to non-Hermitian quasiperiodic potential, while the blue circles represent the disorder free lattice sites. Nearest-neighbor hopping (solid line) along each chain is denoted by $t_\text{A/B}$, and the inter-chain hopping as $t_r$ (dotted lines).}
\label{fig:fig1}
\end{figure}

\section{Model and approach}
\label{sec:model}

The system of coupled one dimensional quasiperiodic non-Hermitian chain and a clean Hermitian chain (see Fig.~\ref{fig:fig1}) can be described by the Hamiltonian
\begin{align}
\mathcal{H} &= -\sum_{x \in \{A,B\}} \sum_{j} t_x\left( c^\dagger_{x, j} c_{x, j+1} + \text{H.c.} \right) \nonumber \\
& \quad -t_r \sum_{j} \left( c^\dagger_{A,j} c_{B,j} + \text{H.c.} \right)\nonumber \\
& \quad + \sum_{j} \lambda_B \cos(2 \pi \alpha j + \phi) \, c^\dagger_{B, j} c_{B, j},
\label{eq:Eq1}
\end{align}
where $x \in \{\text{A},\text{B}\}$ labels the chains, the particle creation (annihilation) operators $c^\dagger_{x,j}$ ($c_{x,j}$) act on site $j$ of chain $x$. 
The first term  represents the standard tight-binding hopping between nearest-neighbor sites within each chain, with hopping strength $t_x$.
The second term denotes the inter-chain tunneling term that couples corresponding sites of the two chains via the hopping amplitude $t_r$.  
The third term introduces an on-site quasiperiodic modulation of the form $\cos(2\pi \alpha j + \phi)$ in chain-B. We choose $\alpha = (\sqrt{5} - 1)/2$ the Golden mean ratio which ensures quasiperiodicity in chain-B and assume the global phase shift $\phi=ih$ to introduce non-Hermiticity in the potential. The chain-A is a clean Hermitian lattice. 
We investigate the model numerically under periodic boundary conditions (PBC) by employing exact diagonalization. Throughout this work, we consider a lattice of size $L=610$ sites in each chain and set the intra-chain hopping amplitude $t_\text{A/B}=1$ as the unit of energy, unless specified otherwise.




We explore the system by solving the single-particle eigenvalue equation corresponding to the model shown in Eq. 1 as,  
\begin{equation}
\mathcal{H} \ket{\psi} = E \ket{\psi},
\end{equation}
For a lattice of size \( L \), the spectrum consists of \( 2L \) eigenvalues, \( E = \{ E_j \mid j = 1, \ldots, 2L \} \). Unlike Hermitian systems, where all eigenvalues are guaranteed to be real, non-Hermitian Hamiltonians \( (\mathcal{H} \neq \mathcal{H}^\dagger) \) can host complex spectra. The emergence of such complex eigenenergies not only signals the breaking of conventional spectral symmetries but also directly impacts the localization properties of the corresponding eigenstates.

In the non-Hermitian ladder, the system can be in delocalized, localized or 
in intermediate phases, where the later is known to host both the localized and delocalized states. These 
different regimes can be distinguished using participation ratios. For an 
eigenstate $\ket{\psi_j} = \sum_{x \in \{A, B\}} \sum_{n=1}^{L} 
\psi^j_{x, n} \ket{x, n}$, the inverse participation ratio (IPR) and the 
normalized participation ratio (NPR) are defined as
\begin{align}
&\text{IPR}_j = \sum_{x \in \{A, B\}} \sum_{n=1}^{L} \left| \psi^j_{x, n} \right|^4, \\
&\text{NPR}_j = \left( 2L \sum_{x \in \{A, B\}} \sum_{n=1}^{L} \left| \psi^j_{x, n} \right|^4 \right)^{-1}
.
\end{align}
In the thermodynamic limit, extended states are characterized by 
$\text{IPR}_j \rightarrow 0$ and $\text{NPR}_j \rightarrow 1$, 
whereas localized states satisfy $\text{IPR}_j \rightarrow \xi_j^{-1} > 0$ ($\xi_j$ is the localization length) 
and $\text{NPR}_j \rightarrow 0$. To obtain a comprehensive understanding 
of our system’s properties, we also calculate
 \begin{align}
&\eta = \log_{10} \left({\langle \text{IPR} \rangle} \times {\langle \text{NPR} \rangle} \right),
\end{align}
where $\langle ~\cdot~\rangle$ denotes averaging over all states. A finite value of $\eta$ indicates an intermediate regime. These measures thus provide complementary perspectives on the localization transitions in the non-Hermitian ladder~\cite{ME_IP_sarma2020,Roy_prl_2021}.
\begin{figure*}[t]
\centering
\includegraphics[width=2.0\columnwidth]{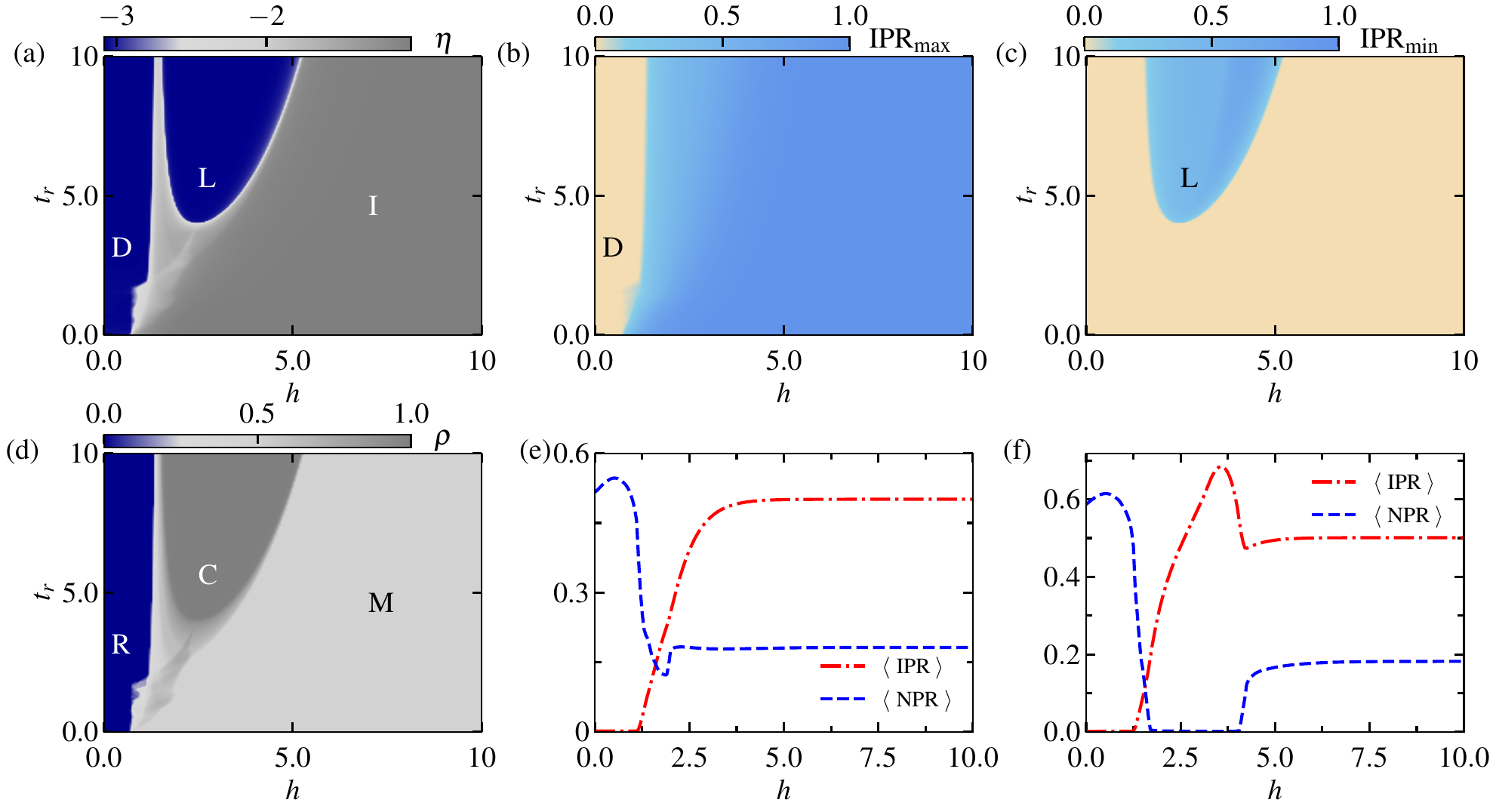}
\caption{
(a) Phase diagram in the \( t_r\text{-}h \) plane shows the delocalized (D), localized (L) and intermediate (I) phases. The intermediate (I, light and dark gray) phase is identified using $\eta$, while the delocalized (D, light brown) and localized (L, light blue) phases are determined from (b) the maximum IPR (IPR\(_\text{max}\)) and (c) the minimum IPR (IPR\(_\text{min}\)). (d) Phase diagram obtained using $\rho$ shows the real (R, blue), complex (C, dark gray), and mixed (M, light gray) regions. Panels (e) and (f) display the variation of $\langle \text{IPR} \rangle$ (red dash-dotted line) and $\langle \text{NPR} \rangle$ (blue dashed line) as functions of the complex phase \( h \) for weak ($t_r=2.0$) and strong ($t_r=6.0$) inter-chain couplings, respectively. The other parameters are fixed at \( t_{A/B} = 1 \), \( \lambda = 1 \), and \( L = 610 \).}

\label{fig:fig2}
\end{figure*}

Beyond such quantitative diagnostics, the spectral properties of the non-Hermitian Hamiltonian themselves also encode signatures of delocalization to localization transition.
When the Hamiltonian \( \mathcal{H} \) possesses $\mathcal{PT}$ symmetry, its spectrum remains entirely real. Upon increasing 
the non-Hermitian parameters, however, the system can exhibit a real-to-complex 
($\mathcal{PT}$-breaking) transition in their spectrum. To characterize the emergence of complex 
eigenvalues, we introduce the density of states with non-vanishing imaginary 
parts,
\begin{equation}
\rho = \frac{N(\text{Im}E \neq 0)}{2L},
\end{equation}
where \( N(\text{Im}E \neq 0) \) counts the eigenstates with a finite imaginary 
component. In the strongly non-Hermitian regime, \( \rho \to 1 \). Thus, the 
localization–delocalization transition and the real–complex spectral transition 
are found to exhibit a one-to-one correspondence. In what follows, we employ 
these measures to analyze the properties of our system.

\section{Results}
\label{sec:results}


In this section, we present our main observations, starting with the trivial case of decoupled leg limit, i.e., by setting $t_r=0$. In this limit, the clean chain (chain-B) is a simple 1D tight-binding chain whose entire spectrum is extended. On the other hand, the chain-A which is subjected to an on-site non-Hermitian quasiperiodic  potential is known to exhibit a delocalization-localization transition at a critical $h$ for any fixed value of $\lambda$.  However, we will show that interesting scenarios appear when the coupling strength between the two chains is made finite. First, we explore the complete phase diagram and the phases and subsequently investigate different limiting situations to explain our findings. Lastly, we complement our numerical results with analytical arguments.

\begin{figure*}[t]
\centering
\includegraphics[width=2.0\columnwidth]{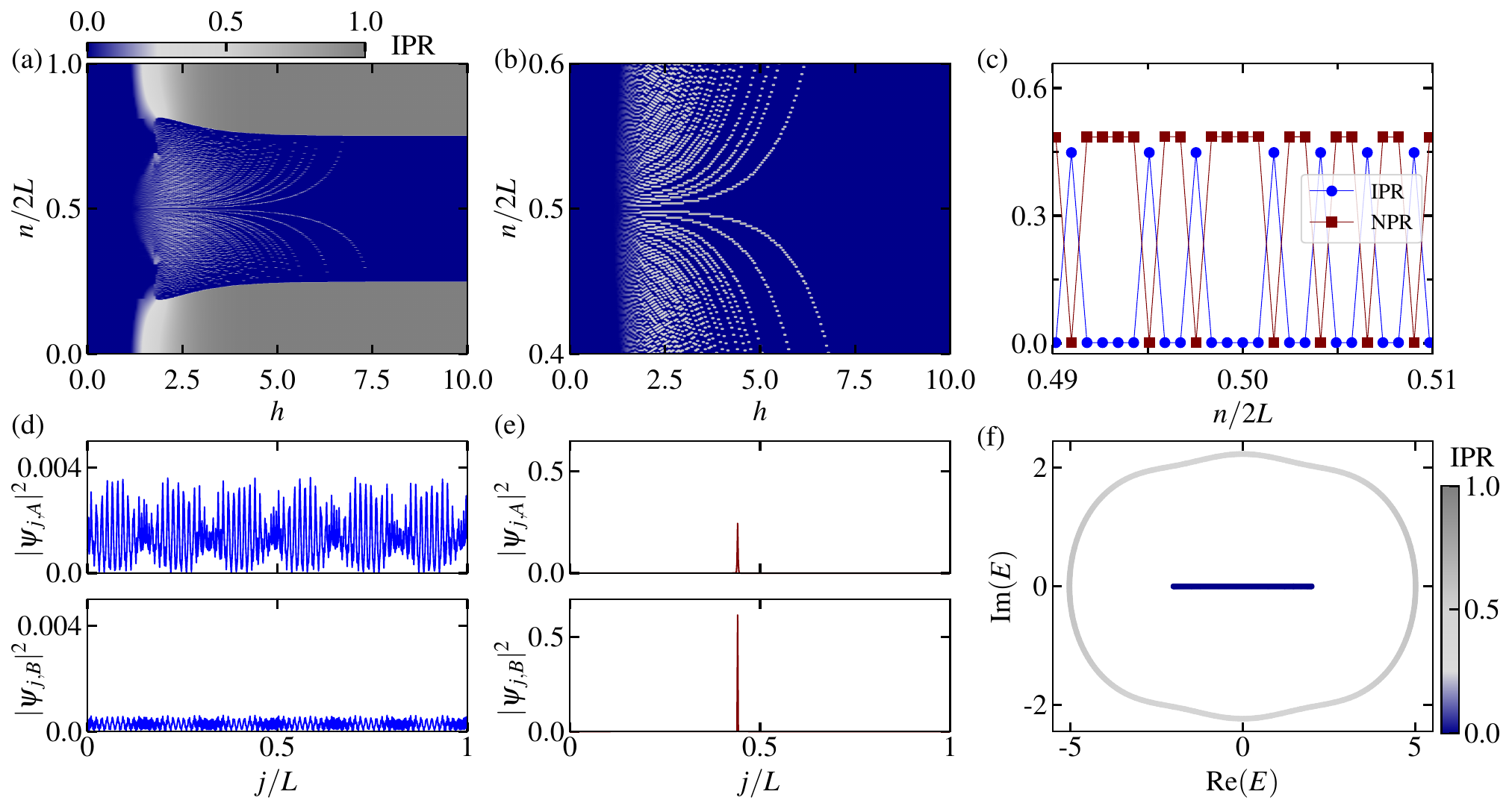}
\caption{(a) Eigenstate index \( (n/2L) \) as a function of \( h \) shows both delocalized and localized states. (b) Enlarged view of central part of (a). (c) A zoomed picture of IPR and NPR of the states with index $n/2L$ from $0.49$ to $0.51$  for \( h = 2.0 \)  depicts the NHCE. The probability density along the site index for both the clean $(|\psi_{j,A}|^2)$ and disordered chains $(|\psi_{j,B}|^2)$, indicating (d) extended state in blue and (e) localized state in red.  (f) Real and imaginary parts of the eigenenergies, along with their corresponding IPR values, are shown for \( h = 2.0 \).  The other parameters are \( t_{A/B} = 1 \), \( \lambda = 1 \), \( t_r = 2 \), and system size \( L = 610 \).}
\label{fig:fig3}
\end{figure*}

\subsection{Phase diagrams}
Before, going to the non-Hermitian model considered in this study, note that in the Hermitian limit of the model~\cite{ME_IP_sarma2020} (i.e., $h=0$ in Eq.~\ref{eq:Eq1}), the system exhibits delocalization to intermediate phase transition as a function of $\lambda$ when the inter-chain coupling $t_r$ is weak. However, for stronger values of $t_r$, the delocalization to localization transition occurs through an intermediate phase~\cite{ME_IP_sarma2020}. In our case, when the non-Hermiticity is introduced in the form of finite $h$, we obtain a phase diagram which is depicted in Fig.~\ref{fig:fig2}(a). The phase diagram is  obtained by plotting $\eta$ as a function of $h$ and $t_r$ for a fixed $\lambda=1.0$. The phase diagram hosts three distinct phases such as the delocalized (D), localized (L) and the intermediate (I) phases. Although, $\eta$ by definition separates the delocalized and localized phases (blue regions) from the intermediate phase (gray region), it fails to distinguish between the delocalized and localized phases (the two blue regions). To clearly identify the localized and delocalized regions in Fig.~\ref{fig:fig2}(a), we plot  IPR$_{max}$ and IPR$_{min}$ which are the maximum and minimum values of IPR in the spectrum, as functions of $t_r$ and $h$ in Fig.~\ref{fig:fig2}(b) and (c), respectively. When the maximum value of IPR ($\text{IPR}_{\max}$) is close to zero, it corresponds to a completely delocalized phase, shown by the light brown region in Fig.~\ref{fig:fig2}(b). 
In contrast, when the minimum value of IPR ($\text{IPR}_{\min}$) stays finite, it 
indicates a completely localized phase, represented by the light blue region in Fig.~\ref{fig:fig2}(c).
Combining these analysis, we identify the D, L and I phases in  Fig.~\ref{fig:fig2}(a). 

From the phase diagram it can be seen that, in the decoupled chain limit ($t_r=0$), the system as a whole is delocalized and then undergoes a transition to an intermediate phase at $h=ln2$. This is because, the chain-B which is a non-Hermitian AA model is known to undergo a D to L transition at this critical $h$ value for the choice of $\lambda=1$~\cite{Longhi_prl_2019}, while chain-A remains completely delocalized. Hence, the entire spectrum together behaves as an intermediate phase. This D-I transition survives up to certain finite values of $t_r$. However, for stronger $t_r$ ($\gtrsim 4.0$), the delocalized spectrum completely turns localized and a D-L transition occurs through an intermediate phase as a function of $h$. This situation is similar to that for the coupled Hermitian quasiperiodic and clean chain discussed in Ref.~\cite{ME_IP_sarma2020}. Interestingly, however, in this limit of $t_r$, if $h$ increases further, then a portion of the spectrum turns delocalized
resulting in an I phase. These transitions can also be seen from the behavior of 
$\langle \text{IPR} \rangle$ and $\langle \text{NPR} \rangle$ of the spectrum. We plot $\langle \text{IPR} \rangle$ (red dot-dashed line) and $\langle \text{NPR} \rangle$ (blue dashed line) as functions of $h$ for two exemplary values of $t_r=2.0$ and $6.0$ in Fig.~\ref{fig:fig2}(e) and (f), respectively. It can be seen that the values of $\langle \text{IPR} \rangle$ ($\langle \text{NPR} \rangle$) are non-zero (zero) in the localized (delocalized) regime and both the quantities are finite in the intermediate regime. 
These phases show a clear one-to-one correspondence with the spectral nature of the states. We can also identify the D, L and I phases from the real (R), complex (C), and mixed (M) nature of their eigenspectrum which are depicted in Fig.~\ref{fig:fig2}(d). The blue region corresponds to purely real eigenvalues, indicating a $\mathcal{PT}$-symmetry preserved phase. In contrast, the light and dark gray regions represent mixed and fully complex spectra, respectively, corresponding to $\mathcal{PT}$ broken region. Notably, the mixed phase in this case aligns well with the intermediate phase in panel (a).  

\begin{figure*}[t]
\centering
\includegraphics[width=2.0\columnwidth]{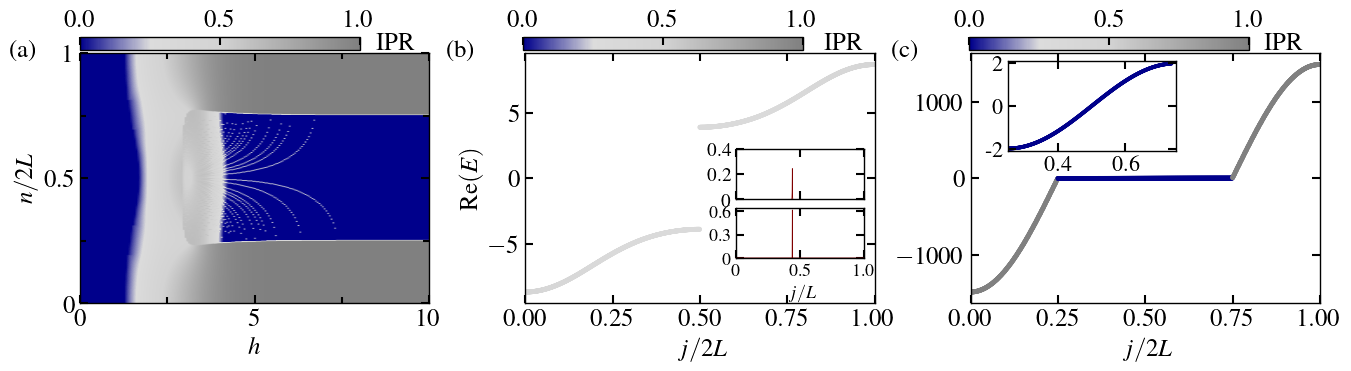}
\caption{(a) IPR of states with respective eigenstate index $(n/2L)$ are plotted as functions of $h$ clearly distinguishes the delocalized, localized, and the intermediate phases. 
(b) Real eigen-energies are shown as a function of the site index, colored according to their corresponding IPR values. The insets show the probability density versus site index, revealing that the states are mainly confined along the rungs, the upper inset corresponds to clean chain, while the lower inset shows for the disordered chain. 
(c) For $h = 10$, shows the real eigen-energies as a function of the site index, the inset zooms into the middle of the spectrum to highlight delocalized features. The parameters used are $t_{\mathrm{A/B}} = 1$, $t_{r} = 6$, $\lambda = 1$, and system size $L = 610$.
}
\label{fig:fig4}
\end{figure*}
 
This analysis clearly distinguishes the D, I, and L phases and also establishes a 
one-to-one correspondence with the spectral phases R, M, and C. Note that the I phase in the large $h$ and large $t_r$ limit also connects to the I phase of the decoupled ($t_r=0$) and weakly coupled  ($t_r \ll t
_x$) chain limit. 
Apart from this we obtain an interesting feature in the spectrum known as the non-Hermitian comb effect (NHCE)~\cite{COMB_2025} both in the weak and strong inter-leg coupling limits which we discuss in the following subsection. 

\subsection{Non-Hermitian comb effect}
To understand the spectral behaviour of the system, we divide the phase diagram into two regimes: (1) weak coupling limit (range of $t_r$ that favours D-I transitions with $h$), (2) strong coupling limit (range of $t_r$ that favours D-I-L-I transition) in the phase diagram shown in Fig.~\ref{fig:fig2}(a). 
To analyze the spectrum of the composite system, we examine the eigenstates of the system for fixed $t_r$ and by varying $h$. To this end, we consider two exemplary values of $t_r$ such as $t_r=2.0$ and $t_r=6.0$, which lie in the weak and strong inter-chain coupling limits, respectively.

In Fig.~\ref{fig:fig3}(a), we show the eigenstates as a function of \( h \), along with their corresponding IPR values for $t_r=2.0$. 
We obtain that for small values of $h$ i.e. $0<h\lesssim  1.25$, all the states are delocalized (blue region on the left). With increase in the value of $h$, localized states start to appear near the edges of the spectrum (gray region). Interestingly, the middle of the spectrum exhibits a coexistence of localized and delocalized states, without a well-defined mobility edge separating them.
An enlarged portion of the spectrum is shown in Fig.~\ref{fig:fig3}(b) for clarity. This behavior differs from the conventional notion of a mobility edge, where a well-defined energy separates the localized state from the delocalized ones in the spectrum~\cite{Mott01011967}. This peculiar behaviour in the spectrum is the signature of the NHCE which can also be regarded as many bound (localized) states in continuum~\cite{BICS_THEORY1, BICS_THEORY2}. The features of NHCE dissapear with increase in $h$ and we obtain an I phase with well separated localized and delocalized states.  

To clearly visualize the comb pattern in the spectrum, we plot IPR (blue circle) and NPR (red squares) of the states according to the eigenstate index \( (n/2L) \) for $h=2.0$ in Fig.~\ref{fig:fig3}(c). We find that delocalized states appear in isolated clusters separated by localized states displaying the features of the NHCE. To confirm the nature of these states, we plot the probability amplitudes of the delocalized (eigenstate index $n/2L = 605/1220$) and localized states (eigenstate index $n/2L = 604/1220$) in Fig.~\ref{fig:fig3}(d) and (e), respectively. The upper and lower panels display the probability amplitudes in the clean $(|\psi_{j,A}|^2)$ and disorder $(|\psi_{j, B}|^2)$ chains, respectively. While the delocalized state clearly shows a spreading of the probability amplitude throughout the lattice, however the localized state exhibits exponential localization. The difference in the amplitude between the chains is due to the effect of quasiperiodic potential in chain-B. 
The wavefunction distribution across both chains indicates that the observed phenomena arise from the composite quantum system as a whole, rather than from the individual chain.
The above analysis demonstrates the NHCE in the systems. 
In Fig.~\ref{fig:fig3}(f), we also show the eigenspectrum in the complex plane for $ h = 2.0 $ with corresponding IPR values of the state. Here, the localized states form a loop surrounding a central cluster of delocalized states. This separation in the complex spectrum provides further evidence for the coexistence of delocalized and localized states.


We also perform similar analysis for strong inter-chain coupling by setting $t_r = 6.0$. Fig.~\ref{fig:fig4}(a)] shows the plot of IPR of the states with eigenstates index $(n/2L)$ as a function of $h$.
Due to the strong coupling between the two legs, the influence of the disordered chain becomes significantly more pronounced in the clean chain. Consequently, the entire delocalized spectrum undergo localization transition through the usual intermediate phase with increase in $h$. Interestingly, in this case, further increase in $h$ leads to a region of spectrum exhibiting NHCE and eventually an intermediate spectrum appears with well separated localized and delocalized states. This behaviour also shows scenario, where some of the already localized states delocalize for larger values of $h$.

\begin{figure}[t]
\centering
\includegraphics[width=1.0\columnwidth]{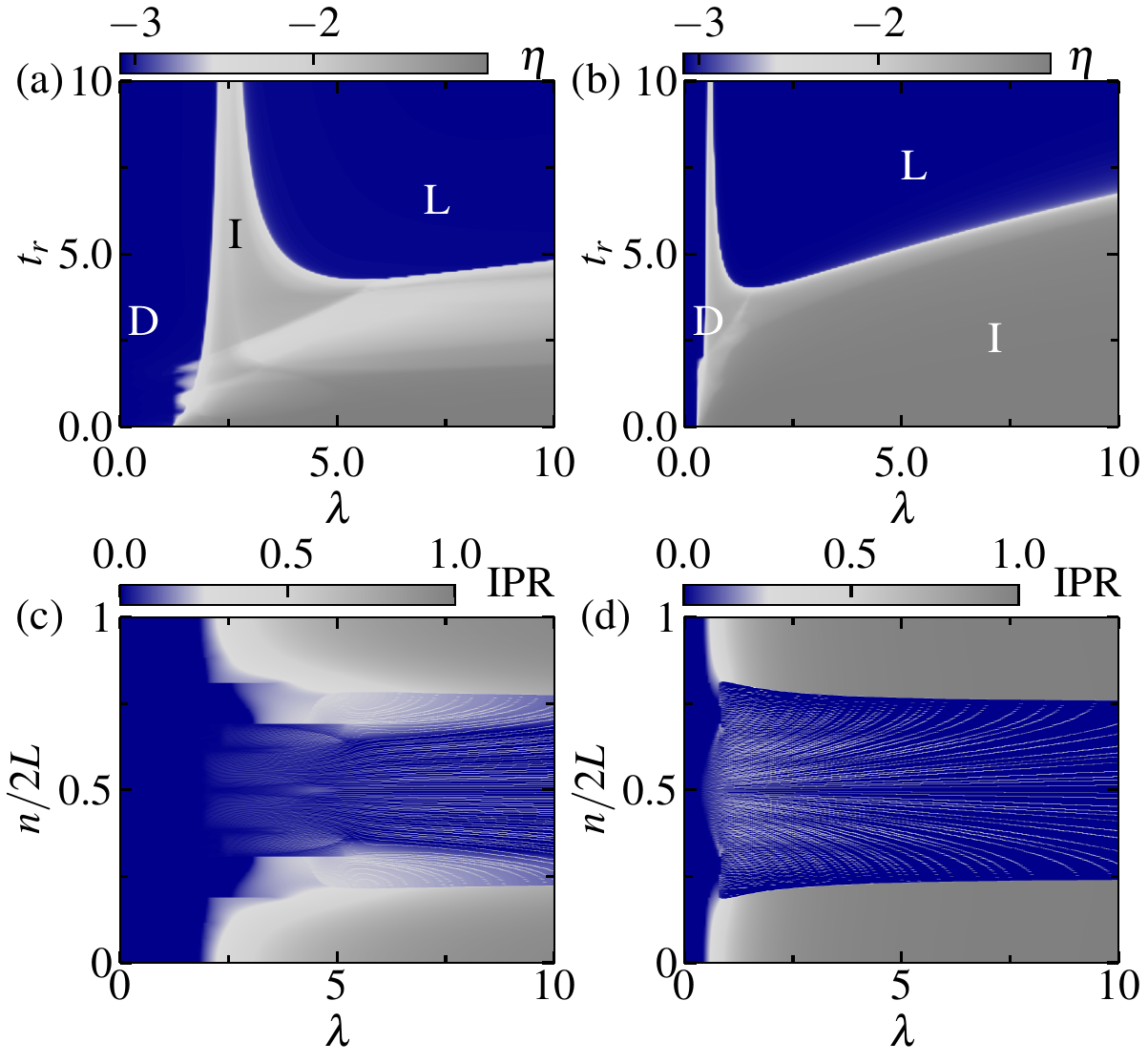}
\caption{Phase diagram in the \( t_r \text{--} \lambda \) plane. Intermediate (I) phase appears in the shaded light and dark gray regions, while the blue areas correspond to the localized (L) and delocalized (D) phases determined from $\eta$ for (a) $h=0.5$ and (b) $h=2.0$. 
IPR of states with respective eigenstate index $(n/2L)$ are plotted as functions of potential strength \( \lambda \) for (c) \( t_r = 1, h=0.5 \), and (d) \( t_r = 1, h=2.0 \). 
The other parameters are \( t_{A/B} = 1 \) and system size \( L = 610 \).
}
\label{fig:fig5}
\end{figure}

To gain further insight into the localized regime, Fig.~\ref{fig:fig4}(b) presents the real part of the eigenenergies as a function of the site index $j/2L$ for $h=3.0$. The color bar indicates the IPR values, clearly showing that all eigenstates are localized. The inset highlights the probability density on both the clean and disordered chains, where in each case the wavefunctions remain confined to the same site index with maximum probability - a phenomenon we refer to as rung localization.
For stronger non-Hermiticity parameter strength, Fig.~\ref{fig:fig4}(c) displays the eigenenergy spectrum at $h=10$. Here, the central region (blue segment) corresponds to the delocalized states with IPR $\approx 0$, accounting for exactly half of the eigenstates, and exhibiting behavior similar to the clean tight-binding chain shown in the inset. In contrast, the states on both sides of the spectrum are colored gray, indicating strong localization. In this limit, the system behaves as two-decoupled chains, which we discuss in detail in a later section.
In the following subsection, we examine cases where the complex phase is fixed while varying the quasiperiodic potential strength $\lambda$, to understand the robustness of the NHCE.

\subsection{Effect of $\lambda$}

We further investigate how the phases evolve with the variation of $\lambda$ for fixed values of $h$. To study this we plot the phase diagram in the $t_r-\lambda$ plane obtained from the values of $\eta$ for $h=0.5$ and $h=2.0$ in  Fig.~\ref{fig:fig5} (a) and (b), respectively. While we obtain similar phase diagrams as obtained with varying $h$, the strength of $\lambda$ tends to favour the localized phases depending upon the value of $h$. For smaller values of $h$, the phase diagram tends to approach the Hermitian case discussed in Ref.~\cite{ME_IP_sarma2020} as expected. 
However, in this situations also we obtain the signatures of NHCE which can be seen from  Fig.~\ref{fig:fig5} (c) and (d) where the states are plotted according to their eigenstates index $n/2L$ and IPR as a function of $\lambda$ for the values of $h$ corresponding to that of the phase diagrams in (a) and (b) respectively. 

The following subsection addresses the origin of NHCE in the coupled system.

\subsection{Limiting case}

\begin{figure}[t]
\centering
\includegraphics[width=0.85\columnwidth]{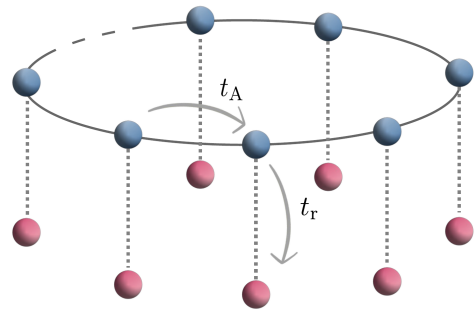}
\caption{
Schematic of the non-Hermitian lattice. Red circles represent sites with a complex quasiperiodic onsite potential, while blue circles denote clean sites. The hopping amplitude along the upper chain is $t_\text{A}$, and the inter-chain (rung) hopping $t_r$ is shown by dashed lines.
}
\label{fig:fig6}
\end{figure}

\begin{figure}[t]
\centering
\includegraphics[width=1.0\columnwidth]{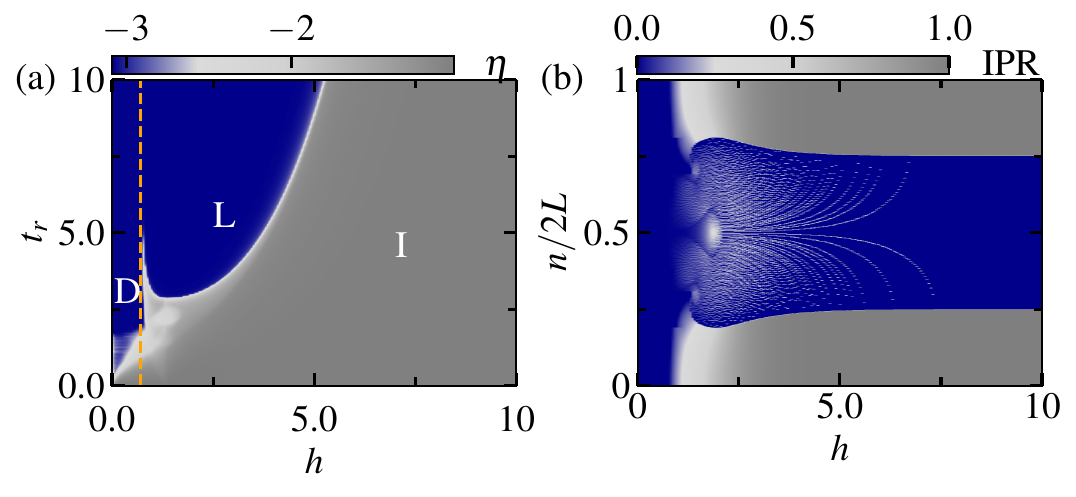}
\caption{(a) Phase diagram in the \( t_r \text{--} h \) plane. Intermediate phase appears in the shaded light and dark gray regions determined by $\eta$, while the blue areas correspond to the localized (L) and delocalized (D) phases. The orange dashed line indicates the critical point $(h=ln2)$ .  
(b) IPR of states with respective eigenstate index $(n/2L)$ are plotted as functions of  \( h \) for weak coupling  \( t_r = 2 \). 
The other parameters are \( t_A = 1 \), \( t_B = 0 \), \( \lambda = 1 \), and system size \( L = 610 \).
}
\label{fig:fig7}
\end{figure}

In this section, we analyze a limiting case of the model defined in Eq.~\ref{eq:Eq1} in order to gain better insight into the scenario discussed above. In particular, we set $t_B = 0$ in Eq.~\ref{eq:Eq1} and argue that this is a minimal model to exhibit NHCE (see Fig.~\ref{fig:fig6}). To demonstrate this we first obtain the phase diagram which is shown in Fig.~\ref{fig:fig7}(a) by plotting $\eta$ as a function of $t_r$ and $h$ while fixing $\lambda=1$. We obtain a qualitatively similar phase diagram  to that is shown in  Fig.~\ref{fig:fig2}(a) corresponding to the original model. However, in the limit of strong $t_r$, the first intermediate region shrinks to a sharp critical line which approaches $h=ln2$ (orrange dashed line)~\cite{Longhi_prl_2019} . Most importantly, this model also possesses the NHCE in the intermediate phase which can be seen from IPR of the states plotted as a function of the eigenstate index and $h$ in Fig.~\ref{fig:fig2}(b). 

To better understand these observations, we complement the numerical results with an analytical calculation. 
We first define the single-particle basis states as,
\[
|\text{A}_j\rangle = c_{\text{A},j}^\dagger |0\rangle, \qquad
|\text{B}_j\rangle = c_{\text{B},j}^\dagger |0\rangle.
\]
With this the eigenstate of the system can be written as
\begin{equation}
|\psi\rangle = \sum_{j=1}^{L} a_j |\text{A}_j\rangle + \sum_{j=1}^{L} b_j |\text{B}_j\rangle.
\label{eq:psi_expansion}
\end{equation}
 $a_j$ and $b_j$ are the corresponding probability amplitudes. 
 Therefore the time-independent Schr\"odinger equations associated with Eq.~\ref{eq:Eq1} can be written as,
\begin{align}
- t_\text{B} b_{j+1} - t_\text{B} b_{j-1} - t_r a_j + \lambda_j b_j &= E b_j, \label{eq:bn_full1} \\
- t_\text{A} a_{j+1} - t_\text{A} a_{j-1} - t_r b_j &= E a_j. \label{eq:an_full1}
\end{align}
where $E$ is the energy corresponding to eigenstate $|\psi\rangle$. Now, let us impose the limiting condition $t_B = 0$,  which is a reasonably good approximation to observe NHCE. In this case, the equation simplifies to,
\begin{equation}
- t_A a_{j+1} - t_A a_{j-1} + \frac{t_r^2}{E - \lambda_j} a_j = E a_j.
\label{eq:an_eff_simple}
\end{equation}

This equation reveals that the on-site potential experienced by the particles moving in the chain-A depends upon the energy of the particle; in other words, the potential landscape gets modified according to the energy of the particle. In this scenario, the following cases arise depending upon the parameter value,
\begin{itemize}
    \item case-1: $t_r\sim t_A$ and $\lambda_j \sim t_A$: In this case, the particles residing on the chain-A have the energy of the order $E\sim t_A$. If $E-\lambda_j\sim \mathcal{O}(1)$ for all the values of $\lambda_j$, then the particle residing on the chain-A experiences a negligible background disorder potential, which is not sufficient to localize the particle. However, if $E-\lambda_j \sim 0$, the potential energy at site-$j$ blows up, trapping the particle at that site, and we observe a localization of the particle. This clearly indicates that if the energy of the particle is in resonance with the on-site potential of the chain-B, we observe localization of the particle in chain-A. This explains the unusual behavior of the intermediate region, exhibiting the comb effect, where localized states appear in an extended continuum when the energy resonates with the quasiperiodic strength, causing the corresponding eigenstates to become localized.
    \item case-2: $t_r \gg t_A$ and $\lambda_j \sim t_r$: From Eq.~\ref{eq:an_eff_simple}, we can see the strength of the disorder potential depends upon $t_p^2$. As we consider $t_r\gg t_A$, the strength of the background disorder potential becomes strong enough to localize the particle irrespective of the energy of the particle. This explains the complete localization of the system in the limit of $t_r \gg t_A$ and $\lambda_j \sim t_r$.
    \item case-3: $\lambda_j\gg t_A$ and $t_r$: In this limit, Eq.~\eqref{eq:an_eff_simple} reduces to the standard tight-binding equation,
    \begin{equation}
    - t_A a_{j+1} - t_A a_{j-1} = E a_j,
    \end{equation}
    which describes a clean chain with nearest-neighbor hopping amplitude $t_A$ and thus the chain-B have no effect upon this.
    The eigenvalue spectrum for this model lies in the range $E \in [-2t_A, +2t_A]$, which matches well with our numerical results for large $h$, which has also been shown in the inset of Fig.~\ref{fig:fig4}(d) for $t_B=1$. This confirms that in this limit, chain-A behaves like an isolated tight-binding chain. This case can also be understood in the following way: when a very large energy offset is created between the two chains, the hopping between the two chains is not allowed due to the energy mismatch. Therefore, due to the large energy mismatch between the two chains, they effectively decouple, and each chain behaves independently.
\end{itemize}

\section{Conclusion}
\label{sec:conc}
In summary, we have explored the single particle spectrum of a non-Hermitian quasiperiodic chain coupled to a clean chain through inter-chain hoppings. By analysing the competing effects of inter-chain coupling and the non-Hermiticity parameter we obtained a phase diagram exhibiting delocalized-intermediate and delocalized-intermediate-localized-intermediate phase transitions in the weak and strong coupling limits, respectively. We also showed that the intermediate phase in a certain situation is nonstandard one which exhibits the non-Hermitian comb effect. We argued that the NHCE in the spectrum is rooted in the decoupled site limit of the quasiperiodic chain. 

Our study presents a model that hosts the NHCE in a system of two coupled chains. This finding suggests the richness of the non-Hermitian systems over their Hermitian counterparts, providing a versatile platform for exploring novel quantum phenomena in the context of localization transition. Our findings also hint at the realization of such NHCE in other lattice geometries, including interacting systems. Given the recent progress in the field of quantum simulation of non-Hermitian systems on platforms such as ultracold atoms in optical lattices and electric circuits, our findings can in principle be experimentally observed.

\section{Acknowledgement}
T.M. acknowledges support from Science and Engineering Research Board (SERB), Govt. of India, through project No. MTR/2022/000382 and STR/2022/000023.

\bibliography{ref}
\end{document}